\documentclass[onecolumn]{revtex4}
\usepackage{graphicx}% Include figure files
\usepackage{dcolumn}% Align table columns on decimal point
\usepackage{bm}% bold math
\usepackage{amsfonts}%\nofile
\begin{document}
\title{Numerical simulation of the internal plasma dynamics \\ of post-flare loops}
\author{Fern\'andez, C.A.,  ultritas@yahoo.com.ar}
\affiliation{Facultad de Ciencias Exactas, F\'\i sicas  y Naturales (UNC-Universidad Nacional de C\'ordoba, Argentina)}
%\title{Numerical simulation of the internal plasma dynamics \\ of post-flare loops}
\author{Costa, A., acosta@mail.oac.uncor.edu}
% \email{acosta@mail.oac.uncor.edu}
\affiliation{Instituto de Astronom\'\i a Te\'orica y Experimental (C\'ordoba, Argentina) \\
Consejo Nacional de Investigaciones Cient\'\i ficas y T\'ecnicas (CONICET) \\ Facultad de Ciencias Exactas, F\'\i sicas  y Naturales (UNC)}
\author{Elaskar, S., selaskar@yahoo.com}
\affiliation{Facultad de Ciencias Exactas, F\'\i sicas  y Naturales (UNC) and CONICET}
\author{Schulz, W., wschulz@efn.uncor.edu}
\affiliation{Facultad de Ciencias Exactas, F\'\i sicas  y Naturales (UNC)}
%\usepackage{txfonts}
%\usepackage{fixltx2e}
%\usepackage{caption}
%\usepackage{sidecap}

%\author[Fern\'andez et al.]{C. A. Fern\'andez$^{1}$ and A. Costa$^{1,2,3}$\thanks{Offprint request Andrea Costa E-Mail:acosta@mail.oac.uncor.edu} and S. Elaskar$^{1,3}$ and W. Schulz$^{1}$\\
%$^{1}$ Facultad de Ciencias Exactas, F\'\i sicas  y Naturales (Universidad Nacional de C\'ordoba, Argentina)\\
%$^{2}$ Instituto de Astronom\'\i a Te\'orica y Experimental (C\'ordoba, Argentina)\\
%$^{3}$ Consejo Nacional de Investigaciones Cient\'\i ficas y T\'ecnicas}
%\begin{document}

\begin{abstract}

We integrate the MHD ideal equations of a slender flux tube to simulate the internal plasma dynamics of  coronal post-flare loops.  We study the onset and evolution of the internal plasma  instability to compare with observations and to gain insight into physical processes and characteristic parameters associated with flaring events. The numerical approach uses a finite-volume Harten-Yee TVD scheme to integrate the 1D$\frac{1}{2}$ MHD equations specially designed to capture supersonic flow discontinuities.  We could reproduce the observational sliding down and upwardly
propagating of brightening features   along magnetic threads of an event occurred on October 1st, 2001. We show that   high--speed downflow perturbations, usually interpreted as slow magnetoacoustic waves, could be better  interpreted  as  slow magnetoacoustic   shock waves. This result was  obtained considering adiabaticity in  the energy balance equation. However, a time--dependent forcing from the basis is needed to reproduce the reiteration of the  event which resembles observational patterns  -commonly known as quasi--periodic pulsations (QPPs)- which are related with large scale characteristic longitudes of coherence. This result reinforces the interpretation that the QPPs are a response to  the pulsational  flaring activity. 
\end{abstract}

\maketitle

\section{Introduction}

Oscillatory processes of several minutes periods  in stellar coronae are associated with large scale quasi--periodic pulsations (QPPs)
in flares. They were  identified in X--ray  stellar coronal flares (Mitra--Kraev et al. \citealp{mit}) and  were later reported for the Sun by Foullon et al. \cite{fou}. Shorter period QPPs in the radio band are known since the early 70's (Aschwanden  \citealp{ash}).
Also,
 systematic intensity perturbations in post--flare loops could indicate that they are the result of
evaporation/condensation, or rising/falling
plasma flow  cycles, caused by the efficient heating
of the flaring plasma from the chromosphere (De Groof et al. \citealp{deg}). This suggests that energy released by flaring pulsations can   induce oscillatory loop trapped processes. 
Many clear quasi--periodic phenomenon are believed to be associated with  wave responses after 
chromospheric evaporation has taken place as the main initial
matter inflow source for flare loops (Nakariakov et al. \citealp{naka2}). Recently, Sigalotti et al.  \cite{sig} simulated the propagation and damping of  localized impulsive perturbations in coronal loops considering different dissipative mechanisms in the linear limit.
As the authors state there is still no consensus over the actual mechanisms from which the waves originate. Moreover,  the importance of  different damping mechanisms is not fully understand due to the fact that different theoretical scenarios can describe similar observational frames.  In this paper we confine to the ideal MHD approach -allowing  the consideration of  nonlinear waves and shocks which give rise to initially non--dissipative damping mechanisms- to evaluate its capability  to adjust observational scenarios. 

In Borgazzi and Costa \cite{bor} we could describe an observational framework of catastrophic evacuation or high--speed
descending flow picture resembling other authors results e.g.
Kjeldseth-Moe and Brekke \cite{kje}; Schrijver \cite{sch}. An upwardly propagating pattern was also registered associated to other parts of the loop system. Also, the frequent recurrence of longitudinal compressible oscillations (e.g. in Borgazzi and Costa \cite{bor},  a quasi--periodic iteration of the event occurred 3 times in 1 h)  were proposed to be associated with either large wavelength slow magnetoacoustic waves or limit--cycle solutions. 
In Costa and Gonz\'alez \cite{cos1} we founded that   limit cycle solutions associated with flow--based  models are
 convergent with large wavelength wave--based models. Magnetoacoustic modes having wavelength of the order of the loop structure describe a basic oscillation between the  kinetic energy of the parallel plasma flows and the internal energy i.e. the thermal perturbation. The structure of the obtained mode is such that in half of the period the disturbance is always positive and the plasma emerges from the chromosphere while in the other half the plasma is negative and descending from the corona to the chromosphere.

In observational scenarios the dynamics is complex making it difficult to isolate the 
erratic flaring from the oscillations. Thus, due to the simplifications that are
assumed for simulating physical quantities,  the
theoretical descriptions generally enables us to distinguish the
behaviour of these  components.
Due to the hight Reynolds number of the corona it could be expected that 
a perturbation could give rise to a slow magnetoacoustic wave or a  rising/falling mechanism that iterates before it decays in a non--dissipative way.

 Our
intention is   to
provide a more accurate description of flaring multi--loop systems which exhibit a combined behaviour, sliding down and upwardly propagating appearance,
 to analyze the plasma dynamics interior to the loops and its relation with the flaring event, i.e.  whether the  oscillatory patterns -which could be associated with MHD waves and/or limit--cycle flow solutions in coronal structures- could be associated to internal or external driving mechanisms. 
As internal mechanism we refer to the development of non--dissipative second standing acoustic modes believed to be  responsible for the induction of QPPs. These QPPs can be  understood in terms of acoustic auto--oscillations, similar to the auto--oscillations in an electric--circuit generator (Tsiklauri  et al. \citealp{tsi2}). The other alternative implies that the QPPs are the resulting features of the external forcing produced by the flaring event.

\section{Numerical method}

The basic MHD  equations, the conservation of mass, conservation of momentum, energy balance, the magnetic induction  equation together with a state equation,   written in the so called conservative form result an hyperbolic--parabolic system of equations. The hyperbolic terms represent the convective effects and the parabolic ones the diffusive parts. These last terms are not considered in this paper so we work  with  an ideal MHD hyperbolic system of equations.  We developed a 1D $\frac{1}{2}$ numerical technique that consists of an approximate Riemann solver that allows to evaluate the variables inside each cell by means of the variation of the flows through the contour of the cell. The software that implements this numerical technique, has proven to be robust solving very demanding benchmarks such as magnetohydrodynamic Riemann problems and Hartmann flows. The MHD Riemann problem solutions are compressible non-steady flows;  the Hartmann flows are incompressible and steady-state flows (Elaskar, et al. \citealp{ela3}a; \citealp{ela4}b; Maglione et al. \citealp{mag}). Thus, the technique is seen as a reliable  computational tool for the description of MHD flows.

Due to its accurate performance we apply   the Harteen--Yee TVD technique (Total Variation Diminishing) which is specially designed to capture discontinuities when dealing with  supersonic flows 
and to achieve a second order approach where the solution is smooth (Yee et al. \citealp{yee}).  Thus, the one non--dimensional equations considering non--stationary flow and written in conservative form are expressed as:
\begin{equation}
\mathbf{U}_{t}+\mathbf{F}(\mathbf{U})_{x}=0\label{1}
\end{equation}
where $\mathbf{U}$ is the state vector of variables
\begin{equation}
\mathbf{U}=(\rho,\rho u_{x}, \rho u_{y}, \rho u_{z},B_{x},B_{y},B_{z},e)^{T}\label{2}
\end{equation}
and the hyperbolic fluxes are 
$$\mathbf{F}= \left( \rho u_{x},\rho u_{x}^{2}-B_{x}^{2}+p+\frac{1}{2}B^{2},\rho u_{x}  u_{y} - B_{x}B_{y} , \rho u_{x}  u_{z} -B_{x}B_{z} ,\ 0 \ , \right. $$
\begin{equation}
\left.  u_{x}  B_{y} - u_{y}B_{x}, u_{x}  B_{z} - u_{z}B_{x},(e+p+\frac{1}{2}B^{2}) u_{x}-\mathbf{B} \cdot \mathbf{u} B_{x} \right)^{T}\label{3}
\end{equation}
$\rho$ indicates de density; $u_{x},$ $u_{y}$ and $u_{z}$ are the velocity components; $B_{x},$  $B_{y}$ and $B_{z}$ represent the components of the magnetic field vector; $p$
is the pressure and $B^{2}=B_{x}^{2}+B_{y}^{2}+B_{z}^{2}.$  The total non--dimensional energy is 
\begin{equation}
e= \frac{p}{\gamma -1}+\frac{1}{2}\rho (u_{x}^{2}+u_{y}^{2}+u_{z}^{2})+\frac{1}{2}\rho (B_{x}^{2}+B_{y}^{2}+B_{z}^{2})
\label{4}
\end{equation}
where $\gamma$ is the ratio of specific heats.

In a quasi--linear form equation~\ref{1} results
\begin{equation}
\mathbf{U}_{t}+\mathbf{A}_{c}\cdot \mathbf{U}_{x}=0\label{5}
\end{equation}
where $\mathbf{A}_{c}$ represents the matrix associated with the Jacobian fluxes, letter `c' indicates that the derivation is with respect to the conservative state variables. A simpler form of the 
Jacobian fluxes is obtain as a function of the primitive variables
\begin{equation}\mathbf{W}=(\rho, u_{x}, u_{y},  u_{z},B_{x},B_{y},B_{z},p)^{T}\label{6}
\end{equation}
Equation~\ref{5} in primitive variables reads
\begin{equation}
\mathbf{W}_{t}+\mathbf{A}_{p}\cdot \mathbf{W}_{x}=0\label{7}
\end{equation}
with  the transformation rule
\begin{equation}
\mathbf{A}_{p}=\mathbf{W}_{t}+\mathbf{A}_{c}\cdot \mathbf{W}_{x}\label{8}
\end{equation}
The conservative form (eq.~5) is generally used for  numerical purposes. 
They are appropriate to obtain accurate jump conditions at discontinuities and shocks (Leveque  \citealp{lev}) and they ensure that mass, momentum
and energy are conserved.  However, the eigenvalue
and eigenvector manipulation is  simpler in the primitive form (eq.~7), thus we
solve the MHD equations using the conservative form and the eigenvectors are
evaluated by means of the primitive formulation.

%%%%%%%%%%%%%%%%%

The resulting Jacobian flux has a null eigenvalue implying that is not possible to use a Riemann solver and thus an alternative Jacobian flux must be implemented. We used a technique  developed by Powell   \cite{pow} and we obtained the primitive eigenvalues which resulted
$$\lambda_{e}=u_{x};  \ \ \lambda_{a}=u_{x} c_{a}; \ \ \lambda_{f}=u_{x} c_{f}; \ \ \lambda_{s}=u_{x} c_{s}; \ \ \lambda_{d}=u_{x}$$
where $e$ stands for  entropic, $a$ for Alfv\'en, and  $s$ and $f$ for  slow and fast respectively.
The Alfv\'en speed and the fast and slow speeds are respectively
$$c_{a}=\frac{B_{n}}{\sqrt{\rho}}; \ \ \ \ \ \  c_{f,s}^{2}=\frac{1}{2}\left( \frac{\gamma p + B^{2}}{\rho}\pm \sqrt{(\frac{\gamma p + B^{2}}{\rho})^{2}-\frac{\gamma p B_{x}^{2}}{\rho^{2}}}\right) $$
We normalized the eigenvectors following the proposal of Zachary et al. \cite{zar} to avoid vector degenerations.

The explicit TVD scheme can be written as 
\begin{equation}
U_{ij}^{n+1}=U_{ij}^{n}-\frac{\triangle t}{\triangle x}\left( <F_{i+\frac{1}{2};j}^{n}>- <F_{i-\frac{1}{2};j}^{n}>\right)\label{9}
\end{equation}
where the functions that determines the numerical fluxes are defined as 
\begin{equation}
 <F_{i+\frac{1}{2};j}^{n}>=\frac{1}{2}\left( F_{i+1}^{n}+F_{i}^{n}+\left[ \sum_{m}\mathbf{R}_{i+\frac{1}{2}}^{m}\mathbf{\Phi}_{i+\frac{1}{2}}^{m}\right] ^{(n)}\right) \label{10}
\end{equation}
$\mathbf{R}$ is the matrix formed by the right eigenvectors of $\mathbf{A}_{c}.$
The dissipation function is expressed as
\begin{equation}
\mathbf{\Phi}_{i+\frac{1}{2}}^{m}=(g_{i+1}^{m}+g_{i}^{m})-\sigma (\lambda_{i+\frac{1}{2}}^{m}+\gamma_{i+\frac{1}{2}}^{m})\alpha_{i+\frac{1}{2}}^{m} \label{11}
\end{equation}
We also used a ``limiter" function approapiated to optimize the one--dimensional flow (Elaskar et al. \cite{ela1}; Maglione et al  \cite{ela2}). 
\begin{equation}
g_{i}^{m}=sgn(\lambda_{i+\frac{1}{2}}^{m}) \max\left( 0, \min(\sigma_{i+\frac{1}{2}}^{m}\mid \alpha_{i-\frac{1}{2}}^{m}\mid, \sigma_{i-\frac{1}{2}}^{m}\frac{sgn(\lambda_{i+\frac{1}{2}}^{m})}{2}\alpha_{i-\frac{1}{2}}^{m})\right)  \label{12}
\end{equation}
with 
\noindent
$\sigma (z) =\mid z \mid $ if  $ \rightarrow \ \ \mid z \mid \geq \varepsilon$ and $ \frac{1}{2 \varepsilon} (z^{2}+\varepsilon^{2})<\varepsilon$; 
$\frac{1}{ \alpha_{i+\frac{1}{2}}^{m}}(g_{i+1}^{m}-g_{i}^{m})$ if $ \alpha_{i+\frac{1}{2}}^{m}\neq 0 $ and $ 0 $ if $ \alpha_{i+\frac{1}{2}}^{m}=0.$\\
The calculation process becomes simpler using the primitive variables to obtain 
\begin{equation}
\alpha^{m}= L_{p}^{m} \cdot (W_{i+1}-W_{i}) \label{13}
\end{equation}
$L_{p}^{m}$ is the left eigenvector of matrix $\mathbf{A}_{p}$ associated with wave $m.$

\section{The problem}

\begin{figure*}%[t]
 \includegraphics[width=5.5cm]{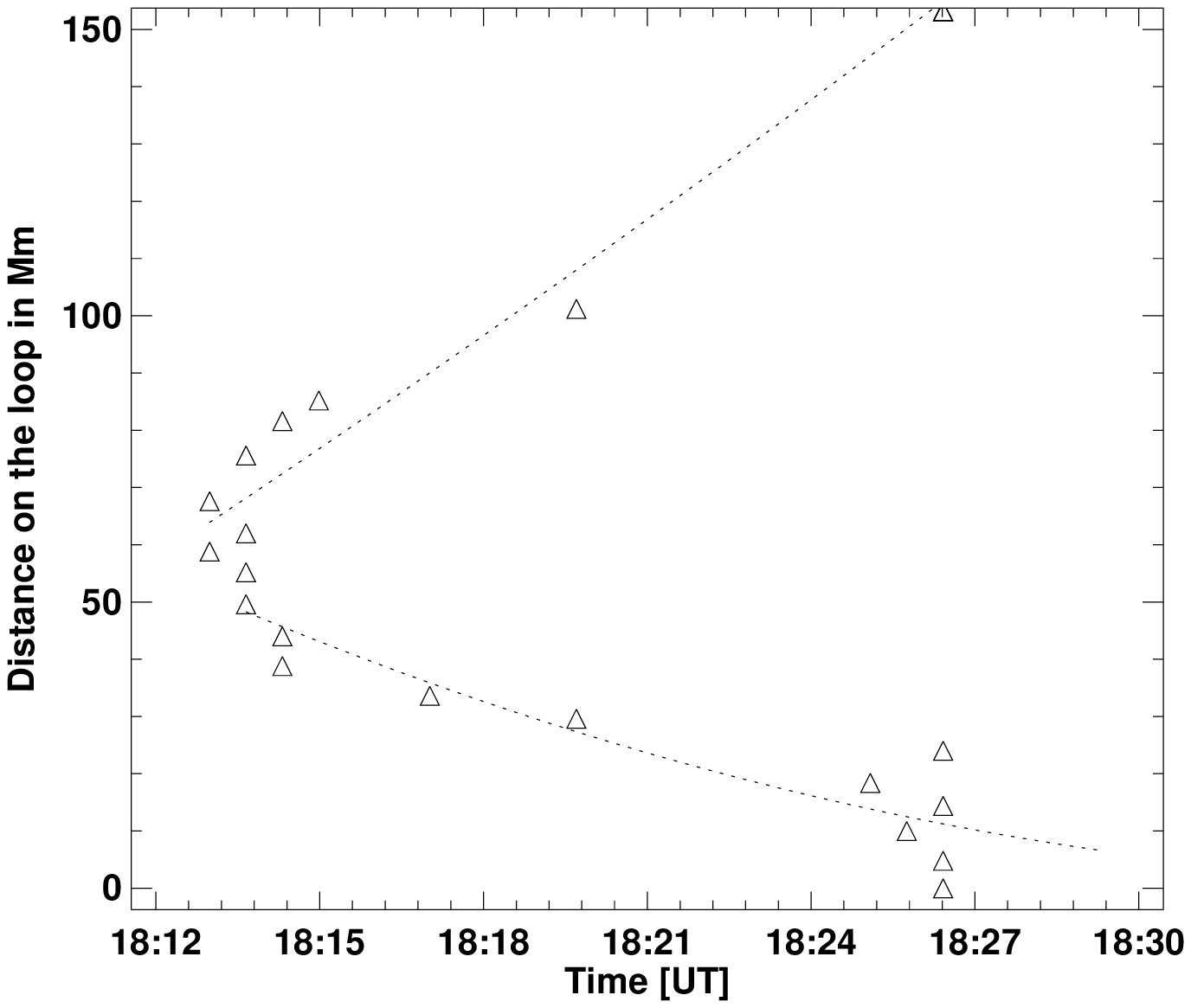}
   \hspace*{74pt}
 \includegraphics[width=5.5cm]{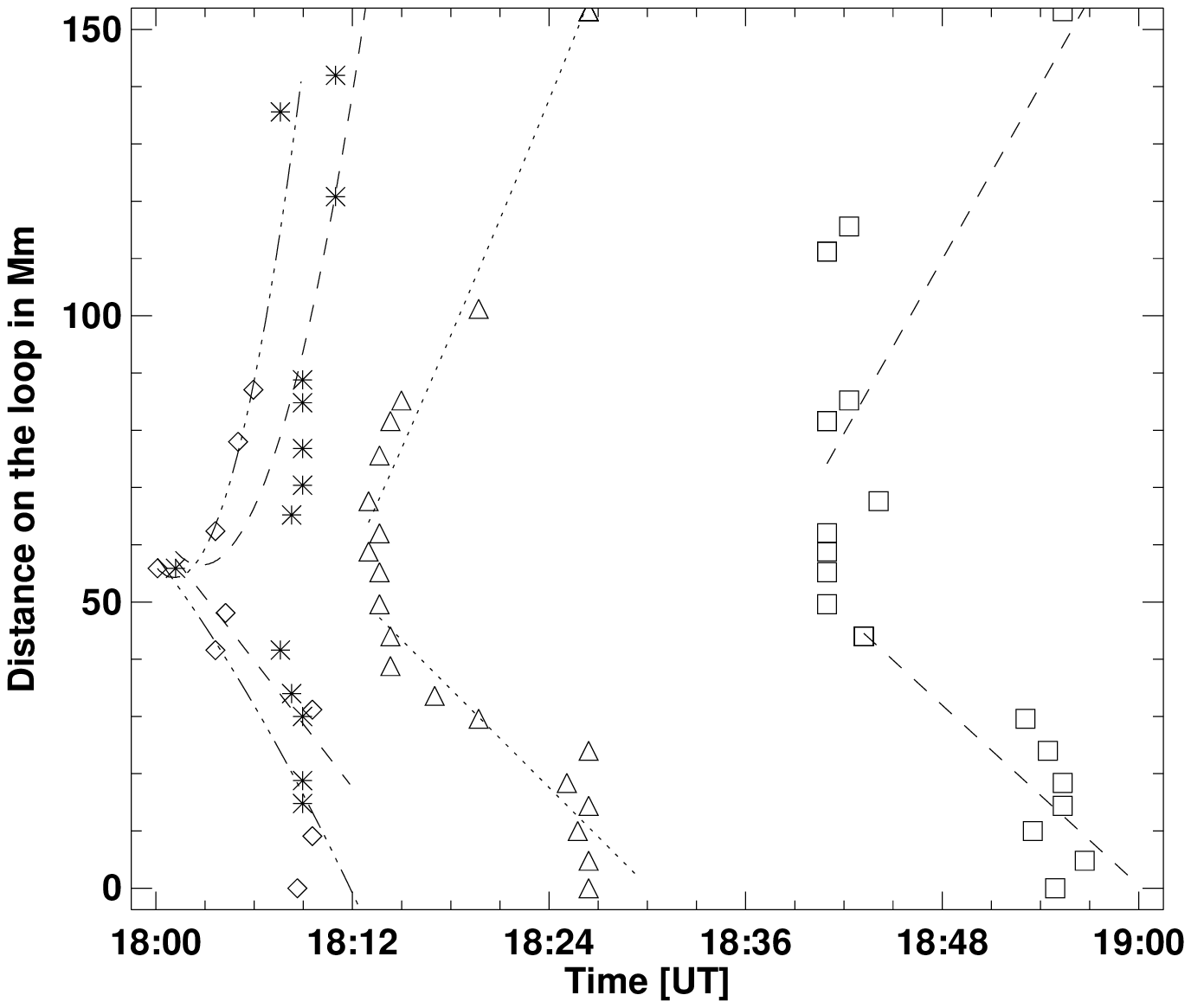}
 \caption{a) TRACE location of the brightening, measured along the loop, as a function of time. The spatial origin is chosen at the loop`s Northern leg. b) Superposition of different events: locations of the brightenings
   as a function of time  measured along a TRACE loop image: $\ast$ for the first TRACE sub--interval; $\bigtriangleup$
     for the
   second TRACE sub--interval and $\Box$ for the third TRACE sub--interval.
   To compare the two telescope results the MICA curve  was superposed ($\diamond$) on the TRACE first
   case. The different  lines of each time sub--interval are the asymptotic curve values.}
   \label{fig:uno}
   \end{figure*}

In Borgazzi and Costa \cite{bor} we studied a post--flare event occurred  October 1st, 2001 which was imagined by both, TRACE space telescope (Transition Region and Coronal Explorer, Handy et al. \citealp{Handy}) and  MICA  land telescope (Mirror Coronagraph for Argentina, Stenborg et al.
\citealp{ste}). We  described an observational framework of catastrophic evacuation or high--speed
descending perturbations which resembles  QPPs  as it was repeated 3 times in 1 h. An upwardly propagating pattern was also registered associated to other parts of the loop system.  Different hypothesis are given in literature to justify why alternatively upward or downward evolving features are not seen (see the discussion in Borgazzi and Costa, \citealp{bor} and references therein). 
Figure~\ref{fig:uno}a shows the location of brightening as a function of time measured over a virtual axis that extends along a whole complex loop i.e. a compound system of neighboring coronal isolated threads which exhibit a coherent behaviour. The origin of the axis was taken at the loop`s Northern foot and the distance was measured along the loop, considering that due to the location of the loop system and to the rough accuracy of the method  projection effects could be discarded.  The figure shows the  sliding down of brightening features from the apex or bifurcation point  towards both legs of the loop.  Figure~\ref{fig:uno}b shows  the superposition of events occurred in the same loop system at different sub--intervals of time. The  coherence is due to the fact that the figure represents the spatial and time behaviour of several apparently isolated tubes. Moreover, as we found that there was a time correlation between relative maxima when comparing different structures of the same active region we suggested that the large longitude chromospheric coherence founded i.e.$\sim 300~Mm$, must be associated with the  forcing from the basis of the whole active region due to the intermittent flaring. However, this is not conclusive, recently Nakariakov et al. \cite{naka2006} developed a model that shows that QPPs observed in a flaring loop can be triggered by MHD oscillations in another loop situated nearby, not necessarily magnetically linked with the flaring one, and thus giving another possible explanation for the coupling of oscillations in an active region.

Solar coronal conditions with large Reynolds numbers are well  fitted by
ideal  MHD plasma models  (i.e.  infinite electrical
conductivity $\sigma \gg 1 $ leading to vanishing viscosity and
ohmic dissipation). We assumed that sources and sinks compensate each other, so the adiabatic energy equation holds,  to  
   investigate if the reiteration of features (Figure~\ref{fig:uno}b) can be associated with the internal dynamics of the plasma, i.e., the second spatial harmonic of the acoustic mode which is believed to be responsible for QPP associated with flaring events (Tsiklauri  et al. \citealp{tsi2}). These undamped  long wavelength solutions can be also thought of as
 limit cycles. These are nearly
longitudinal magnetoacoustic modes that describe a basic oscillation between parallel plasma kinetic energy and plasma internal energy with a characteristic length of $L/4, \ L$ the length of the loop resulting in the balance between  radiative losses and thermal conductive flux   (Costa and Gonz\'alez \citealp{cos1}). Otherwise, the reiteration of features could be associated to an external forcing from the chromospheric basis. This solution is generally suggested by  the chromospheric  coherence observed in the dynamics of many coronal events (e.g. Borgazzi and Costa \citealp{bor}; Mart\'\i nez et al. \citealp{guad}). Of course, this description is not able to reproduce limit cycle models of the type proposed by Kuin and Martens \cite{mar} which are associated with the generation and absorption of heat in open loop systems, neither the model  proposed by Muller et al. \cite{mul} via the assumption of different   radiative loss functions, not taken into account here.

Firstly our intention is to reproduce an individual observational scheme. And secondly we investigate if the reiteration of  features could be explained by an  internal triggered mechanisms or if it requires an external forcing one. In the frame of this model the role played by  the magnetic field is mostly to guide the plasma flow, so  the simulations are performed varying the  thermodynamical quantities of the loop  in  appropiate
 coronal and chromospheric ranges.  

\begin{figure}%[t]
 \includegraphics[width=9.cm]{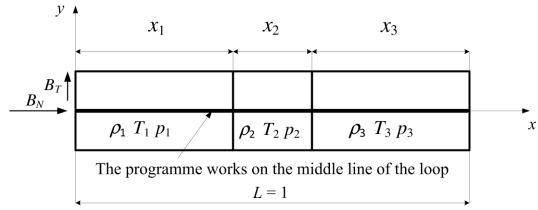}
 \caption{Scheme of the loop description. }
   \label{fig:dos}
   \end{figure}
\begin{figure*}%[t]
 \includegraphics[width=6.5cm]{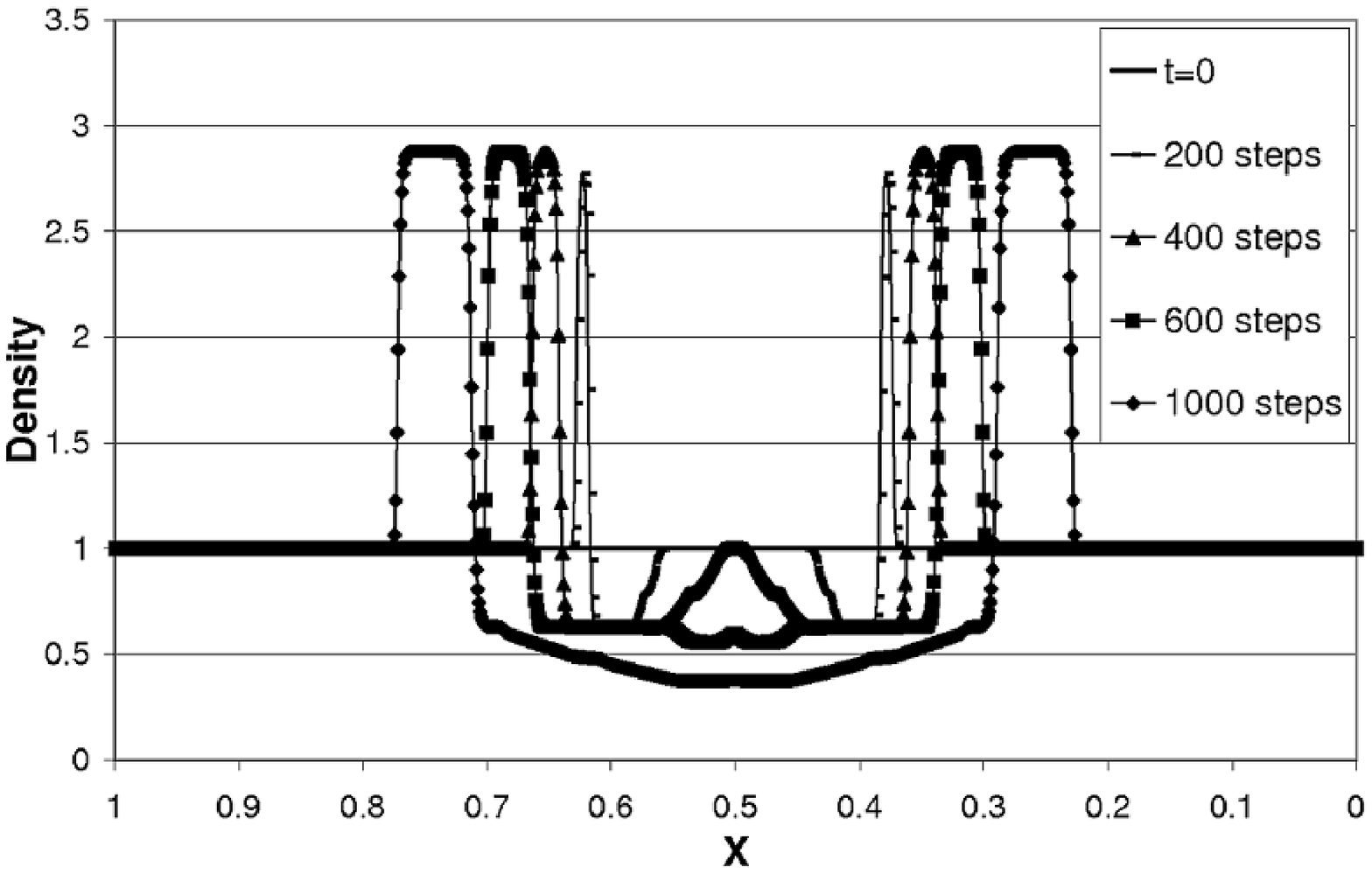}
  \hspace{1.2cm}
  \includegraphics[width=6.5cm]{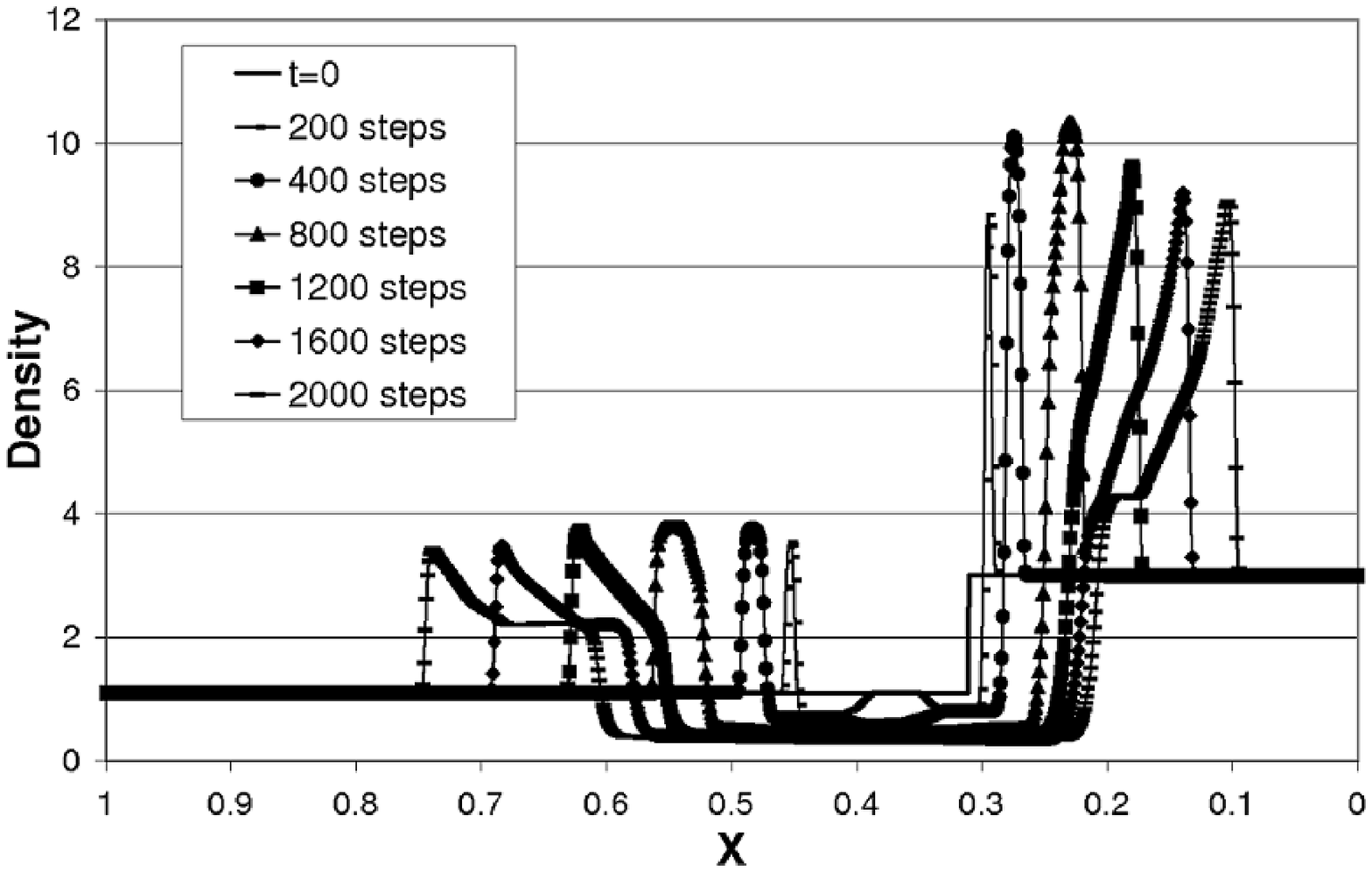}
 \caption{Non--dimensional density -with respect to the background value- as a function of the non--dimensional loop length for different time steps. a) symmetric case, b) asymmetric case. The reference density and longitude values are $\rho_{o}=10^{11} cm^{-3}$ and $L_{o}=150 Mm$ respectively. }
   \label{fig:tres}
   \end{figure*}
\begin{figure}%[
%\hspace{1.0cm}
 \includegraphics[width=3.5cm]{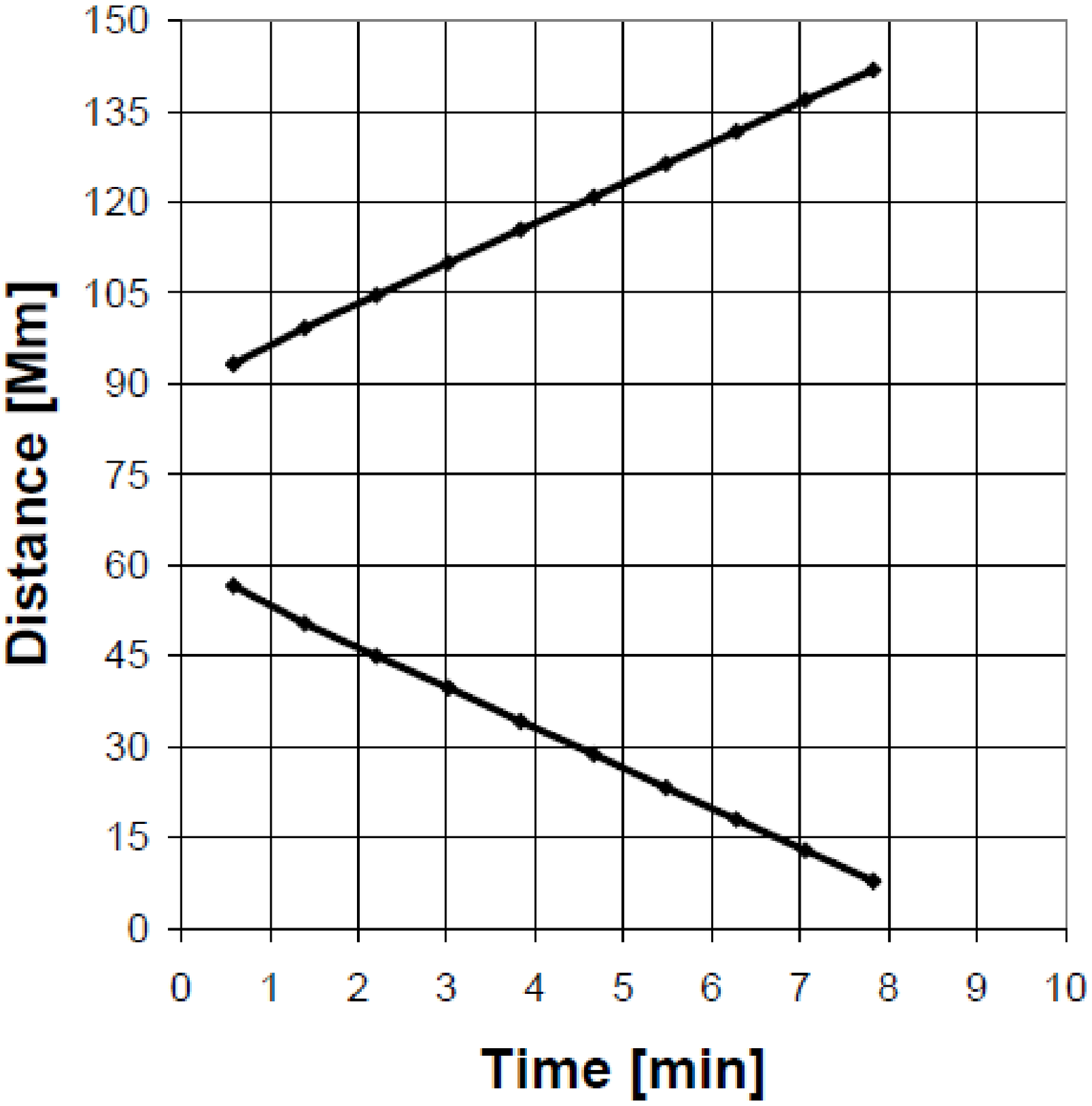}
% \vspace{.5cm}
 \includegraphics[width=5.cm]{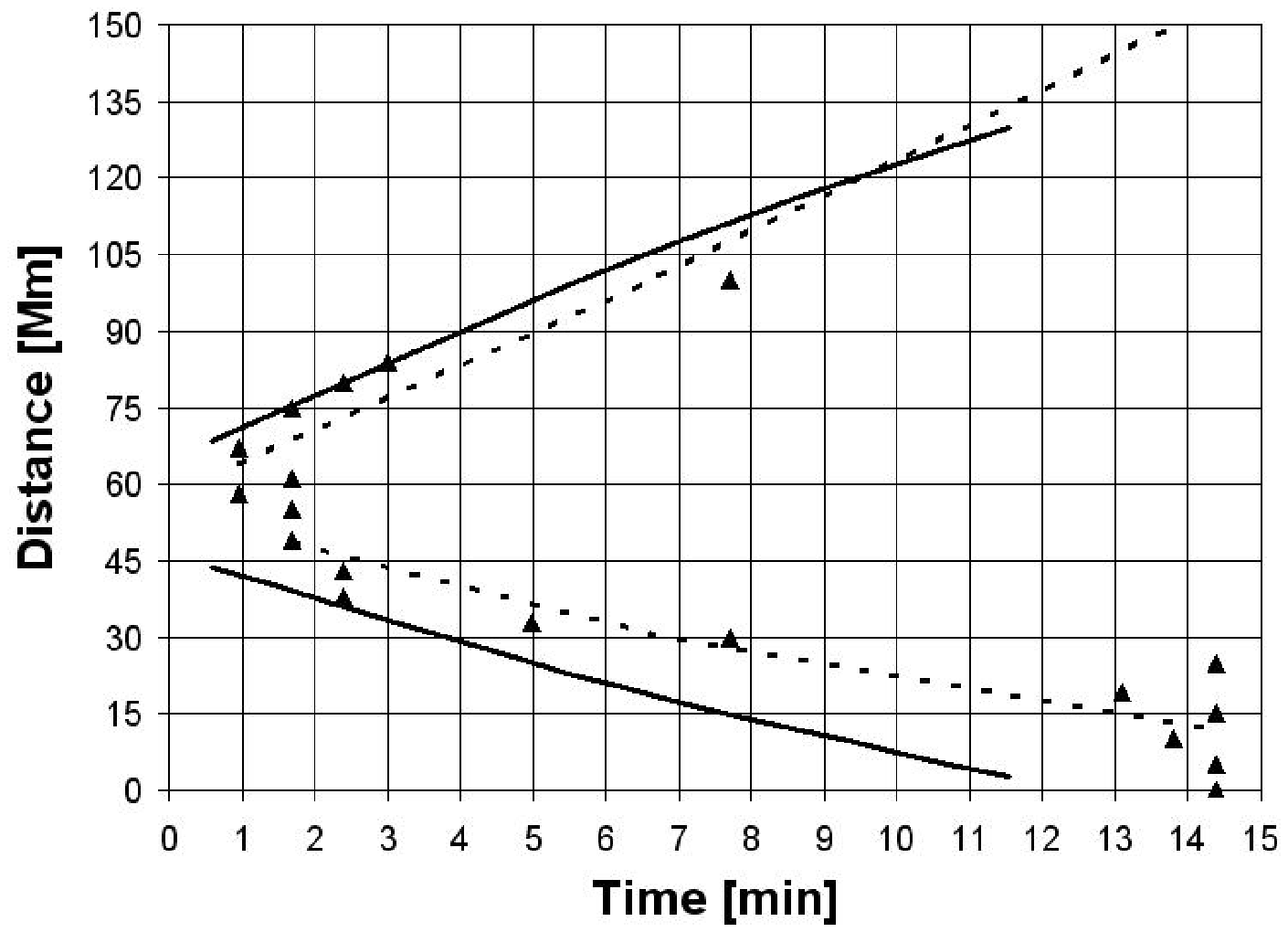}
 \caption{Simulation of the individual a) symmetric case and b) the asymmetric case. The observational data are superimposed .Dot lines correspond to the asymptotic observational values in Figure~\ref{fig:uno}a.  }
   \label{fig:cuatro}
   \end{figure}
\begin{figure}%[
 \includegraphics[width=4.2cm]{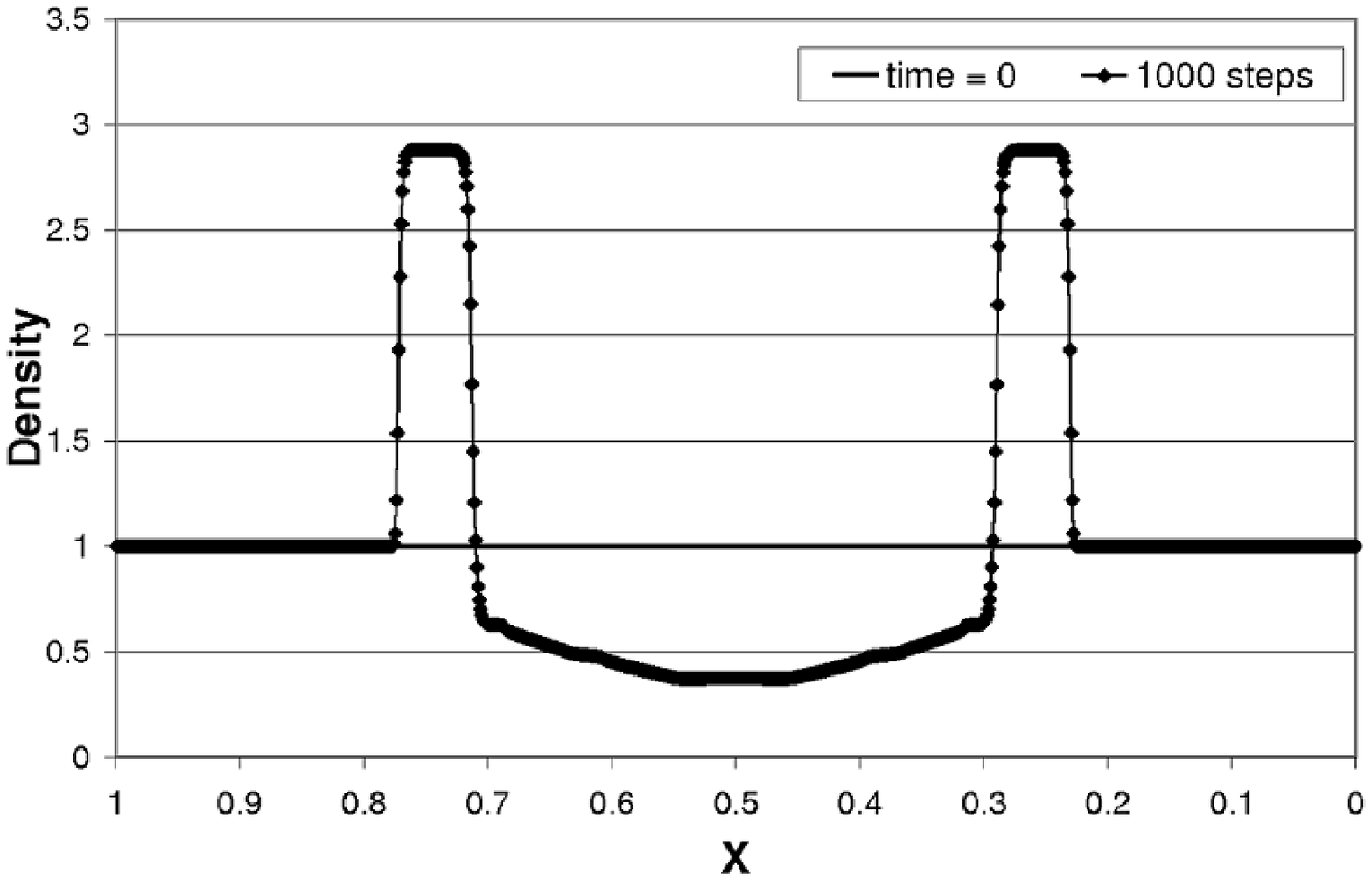}
 \includegraphics[width=4.2cm]{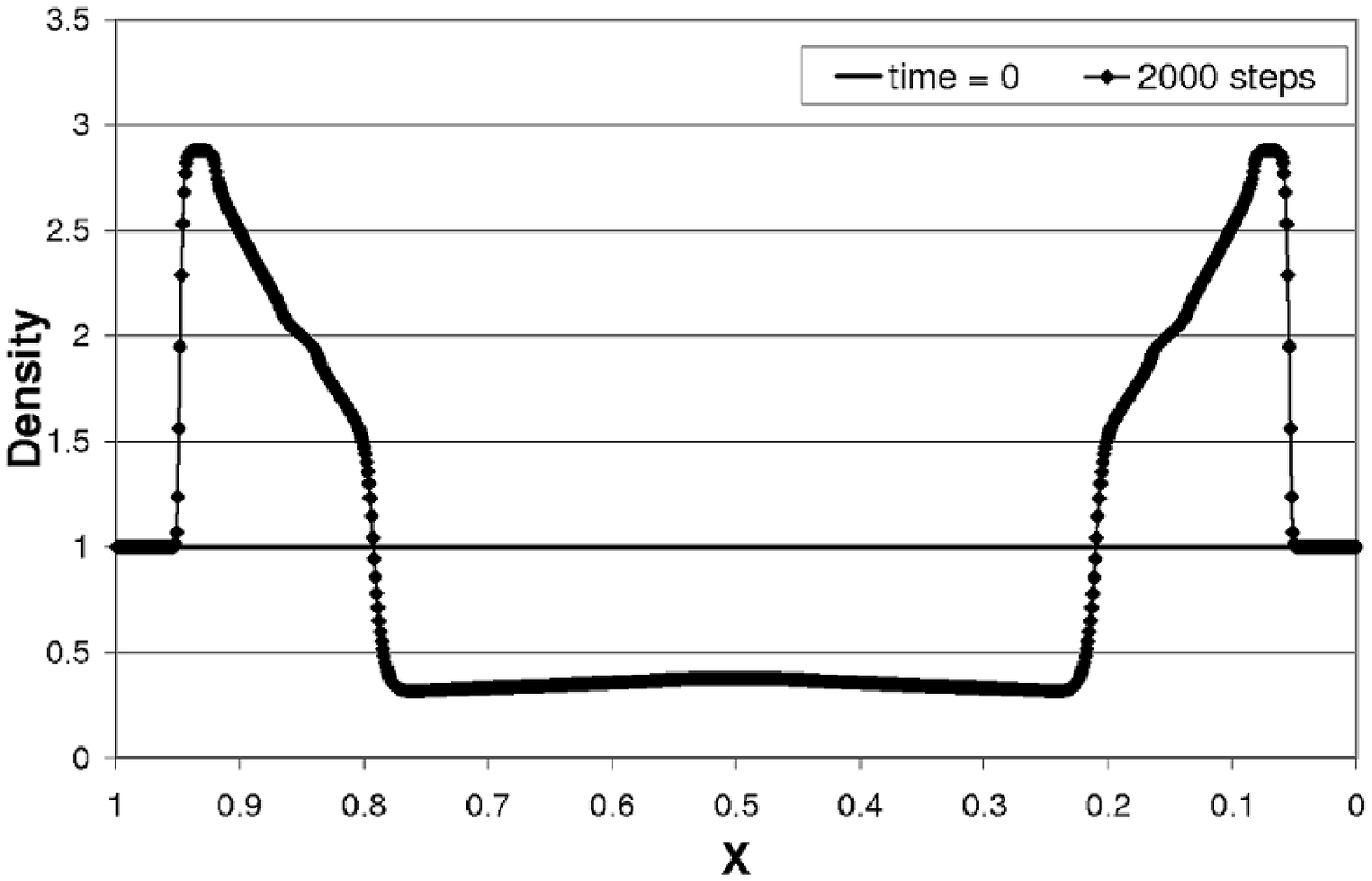}
 \caption{Non--dimensional density as a function of the non--dimensional loop length (same reference values as in Figure~\ref{fig:tres}) for: a) time step $1000$ ($3.84 \ min$). Shock wave fronts $x_{1s}=0.2;x_{2s}=0.8$, contact discontinuities $x_{1c}=0.3;x_{2c}=0.7$, expansion waves $x_{1e}=0.45;x_{2e}=0.55$; b) time step $2000$ ($7.82 \ min$). Shock wave $x_{1s}=0.05;x_{2s}=0.95$, the  expansion waves have interacted with the contact discontinuities and lowered the  density and pressure.}
   \label{fig:cinco}
   \end{figure}

\section{Results and discusion}
\subsection{Simulation of an individual sliding down event }

We represent the   loop system as a straight slender flux tube divided into three  sections $(x_{1},x_{2},x_{3}),  \ \ x_{1}+x_{2}+x_{3}=1$ to integrate the non--dimensional 1D $\frac{1}{2}$ MHD ideal equations. 
The non--dimensional component of the magnetic field along the loop is  $B_{x}=1; \ B_{y}=0.01,$ is an initial transverse perturbation, note that  $B_{x}>>B_{y};$ the reference magnetic field value used is $B_{o}=20 \ G.$   The non--dimensional variables are taken  with respect to the   second section values, e.g. $v_{1}=u_{1}/u_{2}.$ 
Several simulations for different sets of initial conditions were performed. The density values  chosen are $\rho_{i}=(10^{9},10^{10},10^{11}) \ cm^{-3},$  the reference density value used is $\rho_{o}=10^{11}cm^{-3};$ the temperature values are  $T_{i}=(10^{4},10^{5},10^{6},2 \ 10^{6},10^{7}) \ K$ with a reference value of  $T_{o}=(10^{6} \ K).$ The pressure is obtained from the ideal state equation. The reference longitude value is  $L_{o}=150 Mm.$ 
For different  combinations of the state values we obtain the correspondent  evolutions.  Figure~\ref{fig:dos} gives a scheme of the initial conditions. 

\bigskip

\textbf{Symmetric case }

We start with a symmetric partition to study the sensitivity of the system to the initial conditions, i.e. $x_{1}=40\%L_{o},x_{2}=20\%L_{o},x_{3}=40\%L_{o}.$  
Table I displays the six initial condition cases that adjusted the upper slope data of Figure~\ref{fig:uno}a.   They correspond to a temperature step initial condition of the unique non--dimensional case, i.e., all the cases in Table 1. We  describe the spontaneous evolution of the system from this static initial condition; the initial speed value used is  $v_{i}=0.$ Cases I and III can be associated with cold models  and thus short loops. In the frame of this model, the fact that  the case that adjust observations results from initial constant density reinforces the hypothesis of homogeneous density distribution of most loops. Figure~\ref{fig:tres}a shows the evolution of the density front for different  time steps. Figure~\ref{fig:cuatro}a is obtained from Figure~\ref{fig:tres}a assuming that the maximum of the perturbed density is a tracer of the observational features. 
We also obtain 
a resulting transverse magnetic field component  smoothly modulated by the density front (the amplitude of the perturbed  field is  $10\%$ of the non--perturbed one). However, following the classification by Nakariakov and Verwichte  \cite{nak} as the mode does not produce any noticeable perturbation of the loop minor radius it will be considered as a longitudinal mode, i.e., the magnetic field plays the role of being the wave guide of a fundamentally hydrodynamic shock.

A more detailed description  can be obtained from Figure~\ref{fig:cinco}a-b and from  the wave analysis  provided by the numerical  techniques used. The figure displays the  time steps number $1000$ ($3.84 \ $min or $0.16 \ \tau_{o}$  $\tau_{o}$ is the acoustic transit time along half of  the loop) and $2000$ ($7.82 \ $min or $0.32 \ \tau_{o}$) respectively. Time step $0$ corresponds to $\rho(0)=1.$ Figure~\ref{fig:cinco}a shows   two descendant shock wave fronts ($x_{1s}=0.2;  \ x_{2s}=0.8$), two descendant contact discontinuities ($x_{1c}=0.3;  \ x_{2c}=0.7$) and two descendant  expansion waves ($x_{1e}=0.45; \  x_{2e}=0.55$). 
As the shock travels, the temperature, the density and magnetic field  values are increased  making the energy rise to values that could allow the observational detection. Note that the observation of moving  brightening requires that these features must be time and spatially localized and thus, to make the detection possible, the density and temperature must be abruptly diminished again to compound a brightening feature.    This  effect is provided by the  contact discontinuities which is recognized because the pressure and the velocity of the flow are not changed while the wave passes. The expansion waves are recognized because they diminish the  density and pressure  values. From the comparison of time step $1000$  and time step $2000$ (cases $a$ and $b$ of Figure~\ref{fig:cinco}), we see that the expansion wave has a larger speed than  the contact discontinuity, i.e. when the expansion wave reaches the contact discontinuity the  density and the pressure of the whole wave system is diminished. 
From Figure~\ref{fig:tres} we also note that the expansion waves have initially an ascendant phase (compare  the  density distribution at the loop center of steps $200$ and $400$ of the figure); the waves collide at the center and then travel downwards.   
Figure~\ref{fig:cinco}b shows for time step $2000$ the position of the descendant shock  fronts ($x_{1s}=0.05; x_{2s}=0.95$).
Also, the expansion waves have interacted with the contact discontinuities and lowered the density and pressure leaving behind a  coupled nonlinear system of waves, i.e., the features located at $x_{1}=0.2; x_{2}=0.8$ in Figure~\ref{fig:cinco}b can no longer be interpreted as contact discontinuities.

\bigskip

\textbf{Asymmetric case}

To simulate the asymmetry in the location of the bifurcation point of Figure~\ref{fig:uno}a-b we assume 
 $x_{1}=0.31, \ x_{2}=0.12, \ x_{3}=0.57,$  which is in correspondence with the asymmetrical dynamical geometry of the loop system, 
 see Figure~\ref{fig:uno}a. The aim is to reproduce the downward branch slopes of Figure~\ref{fig:uno}a, taking into account that to make the signal detectable the maxima must be higher than the background intensity value, e.g.  Figure~\ref{fig:tres}b. 
Starting from the non-dimensional symmetric case,  the pressure and density of  the segments   $x_{i}$ are changed  to reproduce  the slopes. The slopes result  strongly sensitive to variations of the density, i.e., note from Figure~\ref{fig:tres}b that higher background densities (right line starting at $x=0.3$) imply lower wave speeds (compare the speed of the right and left perturbations). We obtained  few cases that could reproduce the  conditions.
The new non--dimensional initial condition values of segments $(x_{1},x_{2},x_{3})$  that adjusted the observations were  $\rho=(1.1,1.1,3)$  pressure $ p=(0.02,1,0.025)$ and $ T=(0.02,1,0.01)$.  Figure~\ref{fig:cuatro}b is obtained in the same way than in the symmetric case, assuming that the maximum of the perturbed density is a tracer of the observational features. The observational curve is superimposed over the numerical one showing  good
accordance between them. 

In the frame of this first model, it is not possible to give account for the reiteration of the brightening features of Figure~\ref{fig:uno}b. As a consequence of the low density values produced by the pass of the expansion wave, the temperature and density acquire very low values in the back front which are not recovered  in the ascending phase. In all cases, the evolution  from  initial values systematically show a deep  vacuum at the back front, i.e.,  density and temperature diminish to almost $30\%$ of the initial background values. This can also be seen from  Figure~\ref{fig:tres}, the center values are lowered below the background density. However, to investigate if the low density region could be narrowed we  vary the $x_{i}$ partition.  In this case, we obtain that the slope of the curve remained  the same as in  Figure~\ref{fig:cuatro}a and the vacuum region is narrowed but could not be lowered more than $8\%$ of the total length, e.g., for $x_{2}=2 $ the low density region covers  $8 \%$ of the center of the loop. Moreover, the fact that  Figure~\ref{fig:uno}b shows
only descendant features and not ascendant ones could be associated with the instrumental resolution that provides a  limit value to the detection of the  density.  This implies that the density associated with the brightening of all the observed  descendant branches must be higher than the  ascendant ones, and last ones presumably not detectable; this fact  could not be reproduced by  the numerical simulation of this  model. In Costa et al.  \cite{cos} we show that the pass of this type of non--linear waves form a  vacuum tail that can give account of the so--called tadpole phenomenon. 

 \subsection{Simulation of the external forcing from the bases}

\begin{figure*}%[
%\vspace{1.3cm}
 \includegraphics[width=6.5cm]{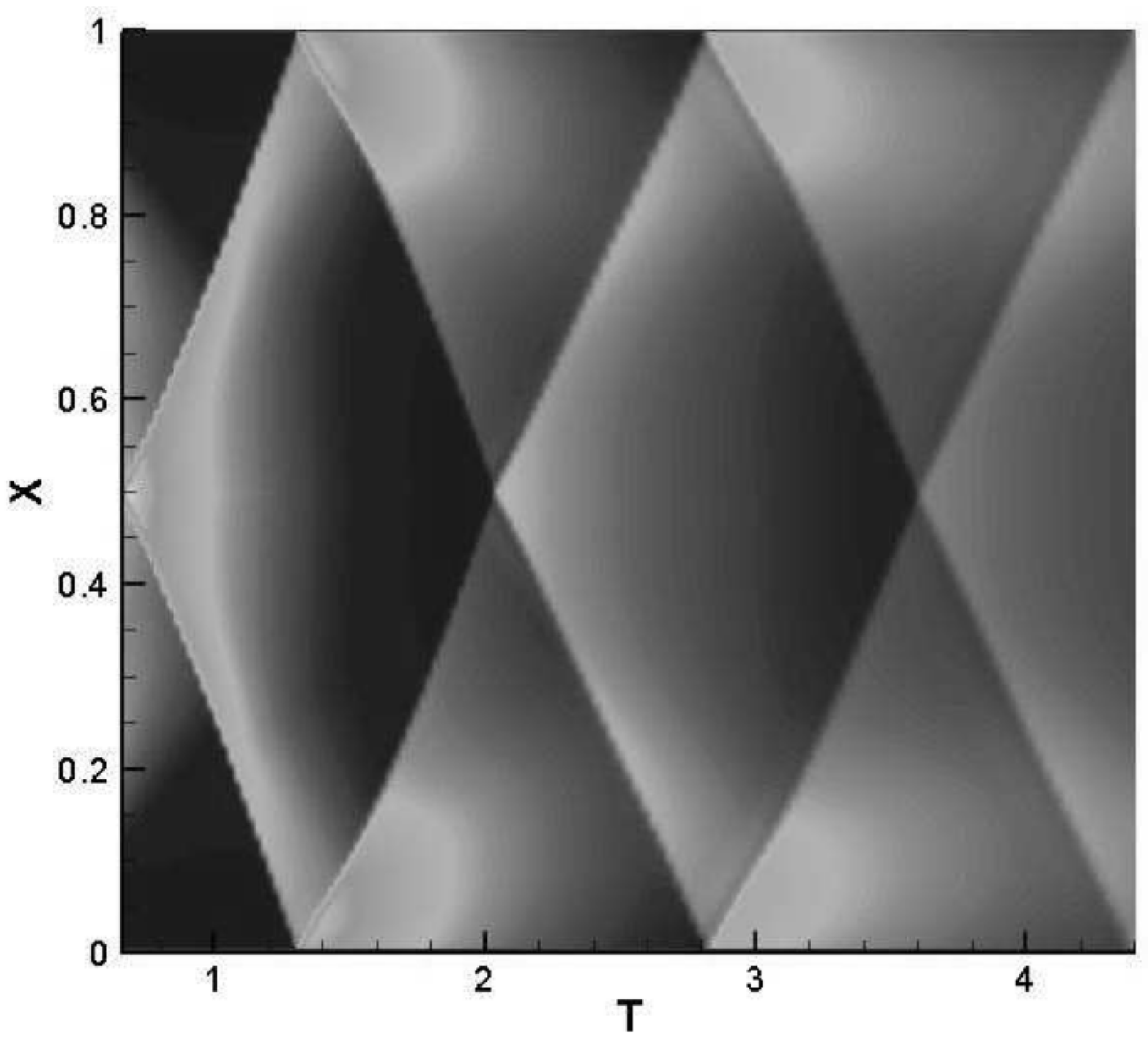}
 \hspace*{1.3cm}
 \includegraphics[width=6.5cm]{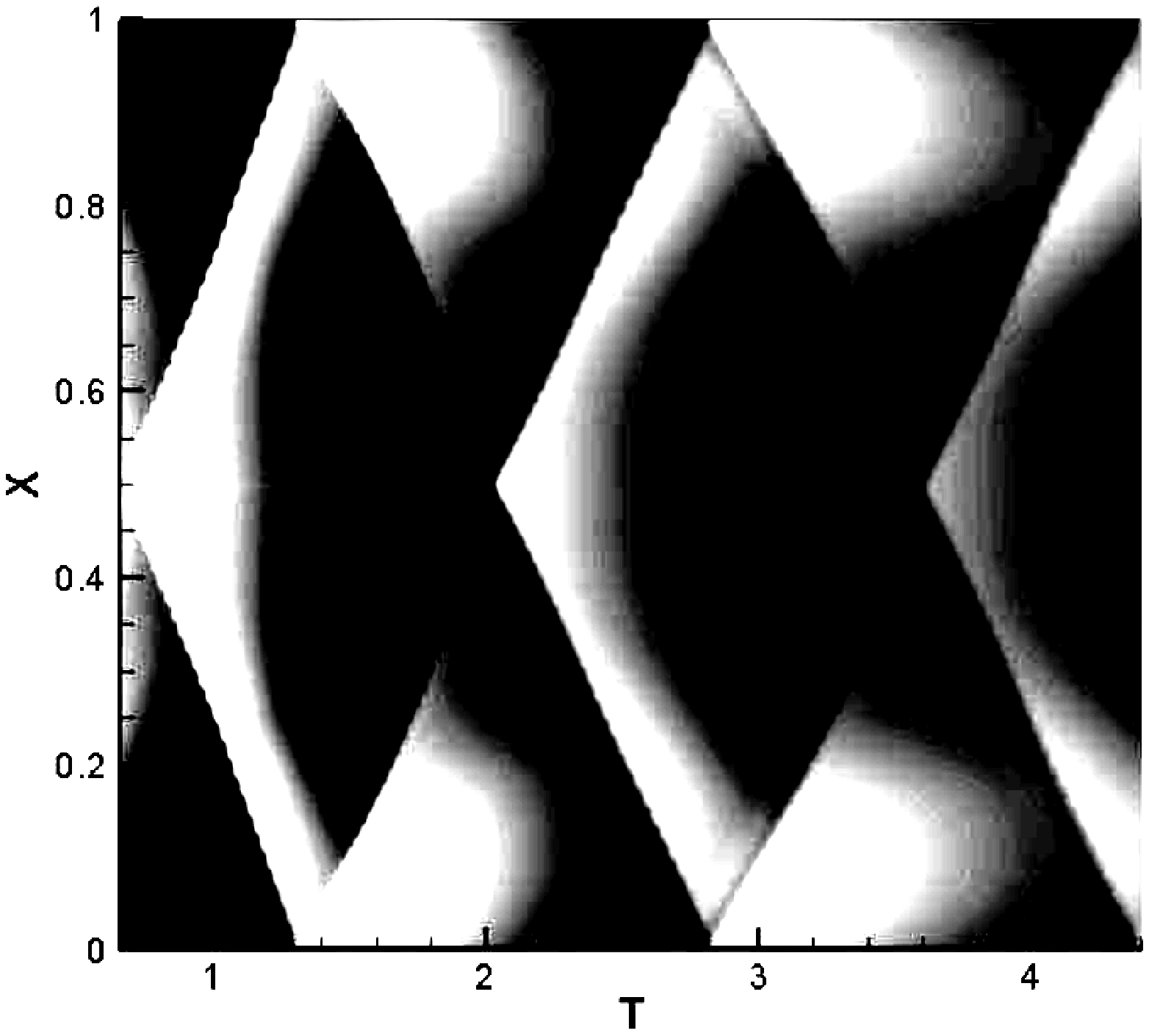}
 \caption{a) Non--dimensional descendant and ascendant density features as a function of the non--dimensional time. b) Non--dimensional descendant density features using the threshold $H.$ $x=1$ is equivalent to $L=150 \ Mm.$  One time division is equivalent to $18 \ min \ 10 sec.$}
   \label{fig:seis}
   \end{figure*}

The different branch speeds of an individual phenomenon were adjusted changing the initial conditions, as in  the first model approach given in subsection~(4.1),   but it was not possible to reproduce its iteration. 
To reproduce the   sliding down features, as in  Figure~\ref{fig:uno}b we decided to investigate the second triggering mechanism proposed. This is the chromospheric forcing from the  loop footpoints produced by a transient oscillation associated with   individual flare burst. The possible forcing functions  are limited by the fact that they must  reproduce the speeds and cadences of  Figure~\ref{fig:uno}b. The temporally and spatially localized impulse deposition, resembling a mass injection in the bases of the loop, was switched off before the start of the simulation. Thus, the forcing must be considered as a boundary condition imposed to  an homogeneous loop, i.e., with initial non--dimensional density $\rho=(1,1,1)$ and temperature $T=(0.1,0.1,0.1).$  Figure~\ref{fig:seis}a shows the density variation  obtained assuming the action of one  individual flare burst modeled 
by a forcing  function $f=A  \ \sin^{6}(\omega t),$ with $  A=3, $ and the frequency $\omega=6.28,$  which corresponds to a deposition of energy of $\sim 10^{3} erg \ cm^{-3}$ on a typical loop.  The value $\omega=6.28$  represents a dimensional characteristic time  of  $\tau=9.3 \ min.$ 
The $x_{i}$ partition  used is the same as in the first symmetric case.

All the descendant branches have  larger values of the perturbed density  than the ascendant ones.  The difference varies between $\sim 20\%$ and $ 40\%$ (for the non--dimensional $x$ value varying between $0.2 $ and $  0.8$) and thus, if the model is accurate, i.e., only the descendant branches are detected,   an intensity threshold $H$ can be numerically obtained. 
In fact, in the EUV and X-ray band, the  emission intensity is modulated by the density perturbations as $H \propto (\rho_{0}+\delta\rho)^{2}$
\begin{equation}
 H_{A}= (\rho_{0}+\delta\rho)_{A}^{2}<H<(\rho_{0}+\delta\rho)_{D}^{2}=H_{D}\label{14}
\end{equation}
and the threshold can be estimated by the above relation.
 $\rho_{0}$ is the background density, $\delta\rho$ the perturbed density, $A$ and $D$ means ascendant and descendant respectively. Taking into account the numerical results, i.e.,  non--dimensional densities $(\rho_{0}+\delta\rho)_{A}=2.45$ and $(\rho_{0}+\delta\rho)_{D}=1.7,$  we obtain a mean value of $<H>=4.446 \ 10^{-22} \ cm^{-6}$ with a $\Delta H=3.113 \ 10^{-22} \ cm^{-6} $ and  $\Delta H/<H>=0.7001. $ Figure~\ref{fig:seis}b is obtained from Figure~\ref{fig:seis}a using the   intensity threshold $\sqrt{<H>}=2.11$ to resemble the observational features, i.e., with only descendant branches.
The ascendant branches are more intense than the descendant ones, however, extended ascendant features ($0<x<0.2$ and $0.8<x<1$) can be distinguished at the location of chromospheric feet in Figure~\ref{fig:seis}b. These features are  correlated with   the  accumulation of observational  points   at the chromospheric footpoints in Figure~\ref{fig:uno}b  and could be indicating the wave front bouncing.

The dimensional speeds  of the numerical descendant branches are: $v_{1}=106 \ km \ sec^{-1},$ $v_{2}=87 \ km \ sec^{-1},$ and $v_{3}=86 \ km \ sec^{-1},$ respectively.  To compare with observation we take the mean value of the speeds of the  northern and southern descendant branches of each event of  Figure~\ref{fig:uno}b. The mean observational speeds are:  $u_{1}=129 \ km \ sec^{-1},$ $u_{2}=82 \ km \ sec^{-1},$ and $u_{3}=65 \ km \ sec^{-1},$ respectively. The time separations between  branches for the numerical case are: $T_{1,2} \sim 24 \ min, $  $T_{2,3} \sim 28 \ min, $ and  $T_{3,4} \sim 28 \ min $ and the correspondent observational quantities (Figure~\ref{fig:uno}b) are:  $T_{1,2} \sim 12 \ min, $  $T_{2,3} \sim 26 \ min, $ and  $T_{3,4} \sim 22 \ min .$

\begin{figure*}%[
 \includegraphics[width=5.5cm]{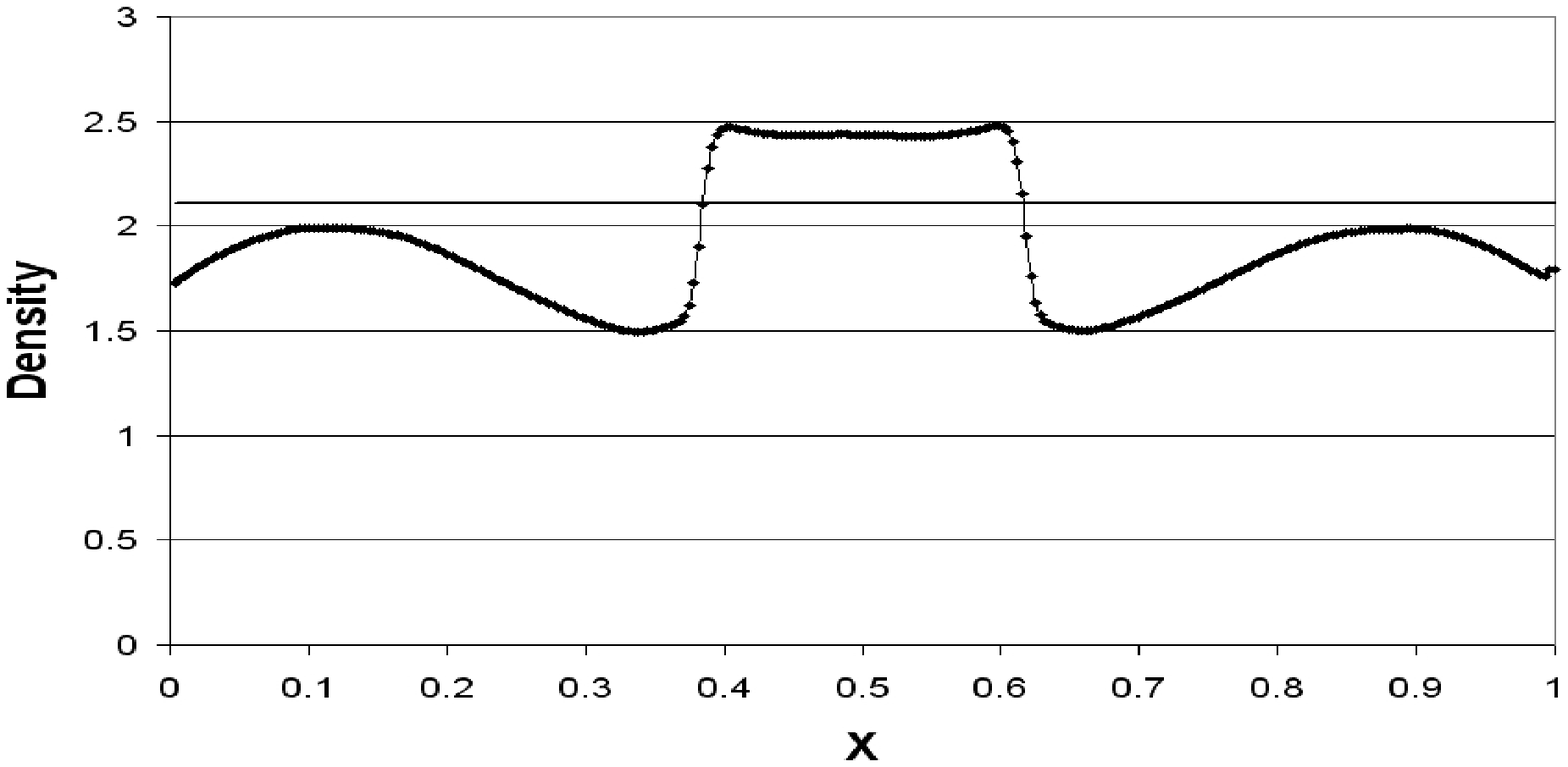}
 \includegraphics[width=5.5cm]{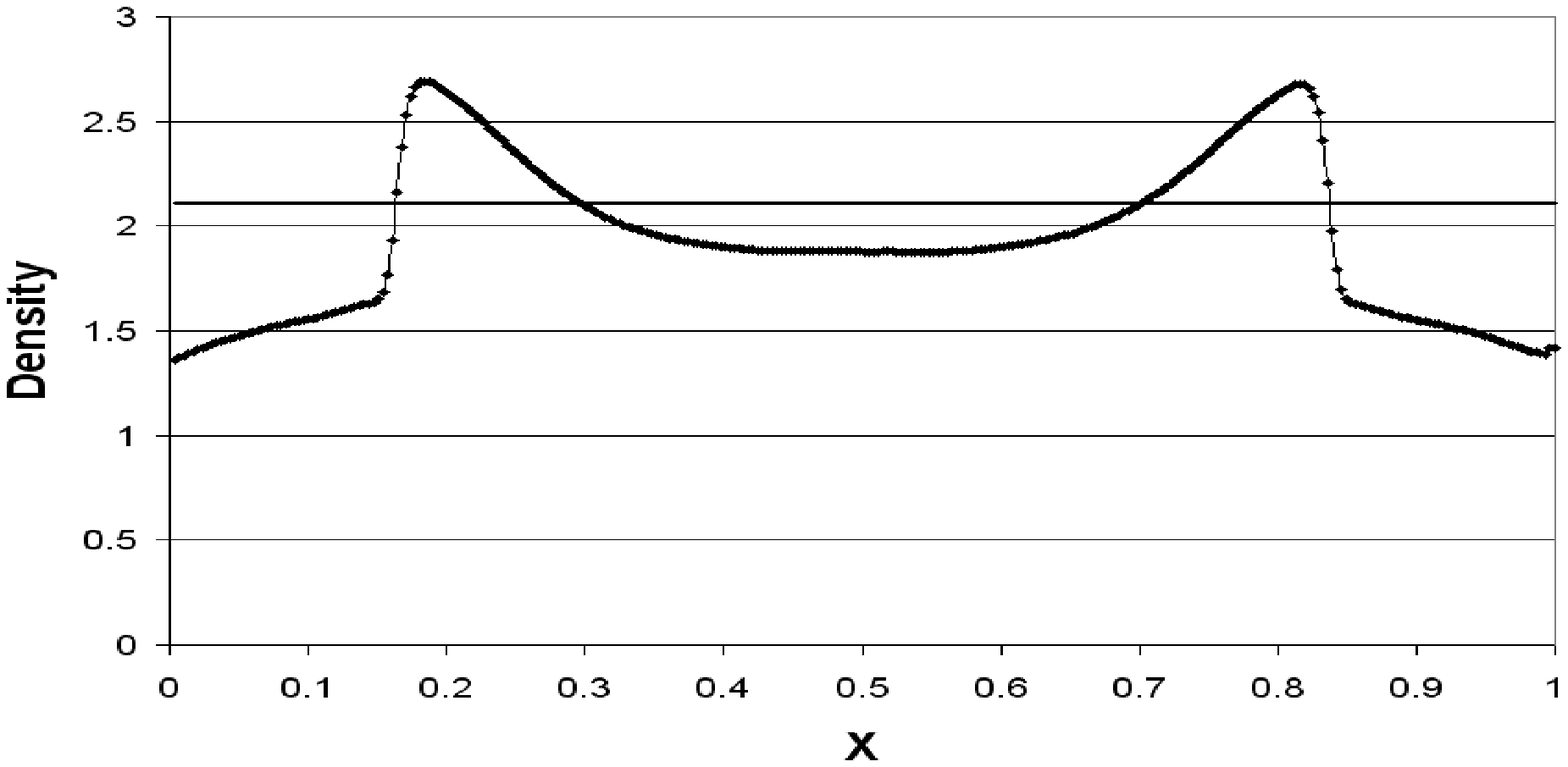}
 \includegraphics[width=5.5cm]{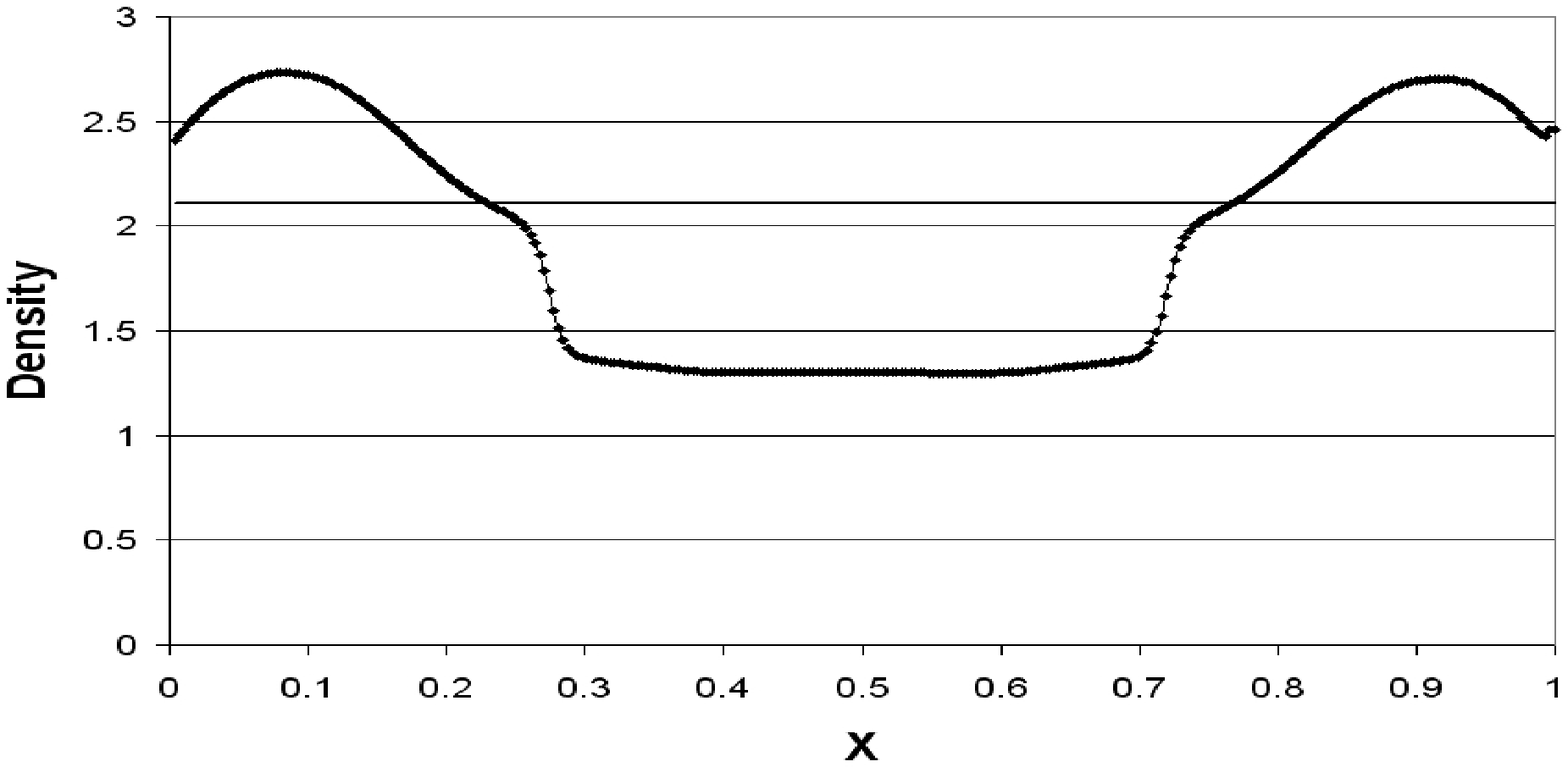}
 \caption{Non--dimensional density as function of the non--dimensional loop length taken from the model in subsection~(4.2) at the a)  non--dimensional time  $2.23,$  two descending shock  fronts $x_{1s}=0.4;x_{2s}=0.6$;  ascending  waves, corresponding to the rebound of the first shock fronts, with  maxima at  $x_{1}\sim 0.1;x_{2}\sim 0.9,$ both below the  threshold $\rho_{H}\sim 2.11;$  b) non--dimensional time  $2.57,$  two descending shock  fronts at $x_{1s}\sim 0.2;x_{2s}\sim 0.8$; ascending  waves lower the density at the center of the figure; c) non--dimensional time  $3.30,$ the interaction of waves smooth the profile, only the ascendant features are detected due to the density threshold, $\rho_{H} \sim 2.11.$}
   \label{fig:siete}
   \end{figure*}

To compare the two model approaches, i.e., spontaneous and forced,  we display  the  density profiles obtained for different   fix temporal values of Figure~\ref{fig:seis}:  $t=2.23,$  $t=2.57$ and $t=3.30.$   The resulting features are shown in Figure~\ref{fig:siete}.   The density profile at $t=2.23$ is shown in  Figure~\ref{fig:siete}a. We see two descending shock fronts at the center of the figure, i.e., $x_{1s}=0.4$ and $x_{2s}=0.6,$ together with  ascending  waves  corresponding to the rebound of the first shock. This last features have their maximum at $x_{1}\sim 0.1$ and $x_{2}\sim 0.9$ respectively and both are   below the non--dimensional  density threshold corresponding to $\sqrt{H_{nd}}=\rho_{H}\sim 2.11.$   The density profile at $t=2.57$ is shown in  Figure~\ref{fig:siete}b. The  two descending shocks are at $x_{1s}\sim 0.2$ and $x_{2s}\sim 0.8.$ At the center the nonlinear interaction of waves lowers the density.  Only descendant shock features are detected due to the density threshold.  The density profile at $t=3.30$ is shown in  Figure~\ref{fig:siete}c.  Again, at the center the density is lowered due to the continue nonlinear interaction of waves.  Only ascendant features from last rebound are detectable due to the density threshold $\rho_{H}=2.11.$ 
 Figure~\ref{fig:siete}b and Figure~\ref{fig:cinco}a have common   sharp edges features  indicating the presence of shock fronts.  From Figure~\ref{fig:siete}c and  Figure~\ref{fig:cinco}b we see common  regions formed by interaction of waves near the bases;  the interaction of waves finish smoothing out  the density profiles.  The differences on the correspondent shapes are related with the differences in the triggering mechanisms (the spontaneous evolution vs. external forcing from the bases), however, the resulting cadences and characteristic speeds are the same as the observational ones. 

Zaitsev and Stepanov  \cite{zai} introduced a model from where they derived relations between flare loop parameters as a function of loop length, period and amplitude of the oscillation: 

$$ T\sim 1.2 \ 10^{-8}\frac{L^{2}}{\tau^{2}}\frac{\Delta H}{<H>} [K]$$
$$n_{e}\sim 2.2 \ 10^{-10}\frac{L^{3}}{Q \tau^{4}}(\frac{\Delta H}{<H>})^{1.5} [cm^{-3}]$$
\begin{equation}
B \sim 6.7 \ 10^{-17}\frac{L^{2.5}}{\sqrt{Q}\tau^{3}}(\frac{\Delta H}{<H>})^{0.75} [G]\label{15}
\end{equation}
where $T$, $n_{e},$ $L,$ $B$ are the temperature, density, length and magnetic strength of the loop. If we take $\tau=9.3 \ min,$ the dimensional characteristic time associated with the flare bursts;  $Q=1,$ the quality factor or the number of individual flare bursts; and the  intensity amplitude of the flare oscillating features as the numerical intensity amplitude, $\Delta H/H,$ we can estimate the temperature, electron density and the magnetic field strength as  
$$T \sim 6.3 \ 10^{6} \ K, \ n_{e} \sim 4.9 \ 10^{9} \ cm^{-3}, B \sim 8.7 \ G.$$ 
To adjust the observations we performed several calculations  varying the  parameters $\Delta H / <H>,$  $Q$ and $\tau$ respectively. We noted that the increasing of $\Delta H / <H>,$ corresponds to more pronounced curves of the descending branches  and that more complex  patterns can be obtained depending on $Q,$  the   number of individual
flare burst chosen  and the time  distance between them, $\tau.$
Taking into account the approximations we assumed and the numerical simplifications of the model we note that the values obtained  are in good agreement with the observational values.  
  The evaluation of the relative importance of the parameters and initial conditions to produce different patterns that could be compared to observations together with 
the consideration of nonideal contributions will be a matter of a new work.

 A few words with respect to the consideration of diffusive terms in the equations must be added. 
Longitudinal modes can be  affected by the thermodynamic properties of the medium. However, the fact that this type of modes is selectively seen in hot loops, where the mode damping is strongest, remains an open question. 
 The scenario we are dealing with shows that nonlinear wave interaction processes  are responsible of  smoothing out sharp edges in a non--dissipative regime. This suggests  that we are describing the stage where a non--dissipative transfer of energy from the large spatial scales to the small ones is predominantly  at work  before the small scales allow an efficient  action of  dissipation.

A novelty of this work is that it proposes that the description in terms of shock waves is more common that what it was supposed. Low  $\beta$ value media have larger values of the Alfv\'en speed than the sound speed. These two characteristic parameters of the coronal plasma determine the type of processes and energies associated to them that can be  involved in its dynamics. The fact that in general, depending on the $\beta$ value, the sound speed is the smallest one suggest that it must not be unusual that compressional perturbations  reach shock wave speeds. 
Inhomogeneous features in the medium can easily  couple the transverse alfvenic component with the longitudinal one and thus transfer energy to the acoustic mode allowing that hydrodynamic shock waves, guided by magnetic fields, can  propagate.


\section{Conclusions}
We have integrated the MHD ideal equations  of a slender flux tube to simulate the internal plasma dynamics of  coronal post-flare loops. 
We could reproduced the observational 
behaviour of brightening features   along magnetic threads of an event occurred on October 1st, 2001, i.e., characteristic speeds and times. 
Our
intention was to    
 analyze the plasma dynamics interior to postflare loops and its relation with  flaring events, i.e.  whether the  oscillatory patterns could be associated to internal or external driving mechanisms. This is, whether
  non--dissipative second standing acoustic modes could be  responsible for the induction of QPPs or if they are  resulting features of an  external forcing produced by the flaring event. We showed that, in the frame of our model, to reproduce the iteration of sliding down observations a time--dependent forcing from the basis is required and thus the interpretation that the  QPPs are responses to the pulsational flaring activity is reinforced.

  We also found that   high--speed downflow perturbations usually interpreted as slow magnetoacoustic waves can be better  interpreted  as  slow magnetoacoustic   shock waves.

\begin{table}
\begin{tabular}{cccccccc}
\hline $Case$&$ \rho_{1}= \rho_{3} \ (cm^{-3}) $&$ \rho_{2} \ (cm^{-3}) $&$ T_{1}=T_{3} (K)  $&$ T_{2} (K)  $&$  wave \ speed  $& $  sound \ speed $  \\  \hline \hline
$I$&$10^{10}$&$10^{10}$&$ 10^{4}$&$10^{5}$&$0.81$&$0.374$
\\ \hline
$II$&$10^{10}$&$10^{10}$&$ 10^{5}$&$10^{6}$&$0.81$&$0.374$
\\ \hline
$III$&$10^{11}$&$10^{11}$&$ 10^{4}$&$10^{5}$&$0.81$&$0.374$
\\ \hline
$IV$&$10^{11}$&$10^{11}$&$ 10^{5}$&$10^{6}$&$0.81$&$0.374$
\\ \hline
$V$&$10^{10}$&$10^{10}$&$ 10^{5}$&$2.10^{6}$&$0.771$&$0.265$
\\ \hline
$VI$&$10^{11}$&$10^{11}$&$ 10^{5}$&$2.10^{6}$&$0.771$&$0.265$ \\ \hline \hline
\end{tabular}
\caption{\label{tab:table2} Values of the six initial condition cases that adjust the observational data of the symmetric case  (see Figure~\ref{fig:uno}a). The wave front speed and the acoustic speed were nondimenzionalized with the Alfv\'en speed $(v_{A}=1)$. }
\end{table}

%%%%%%%%%%%%%%%%%%%%%%%%%%%%%%%%%%%%%%%%%%%%%%%%%%%%%%%%%%%%%%%%%%
\begin{acknowledgements}
We are grateful to Dr. Tamagno for his comments and
suggestions that helped to improve the quality of the paper.
\end{acknowledgements}

\end{document}